\documentclass[aps,pre,twocolumn,showpacs,superscriptaddress]{revtex4}
\usepackage{graphicx}
\usepackage{amsmath}
\usepackage{setspace}
\usepackage{dcolumn}
\usepackage{bm}
\usepackage{epstopdf}
\begin{document}
\title{Dynamical Phase Transitions in TASEP with Two Types of Particles under Periodically Driven Boundary Conditions}

\author{Ay\c{s}e Ferhan Ye\c{s}il}
 \email{yesil@bilkent.edu.tr}
\author{M.~Cemal Yalabik}
\affiliation{Department of Physics, Bilkent University, 06800
Ankara, Turkey}
\date{\today}
\begin{abstract} 
Driven diffusive systems have provided simple models for non-equilibrium
systems with non-trivial structures. Steady state behaviour of these
systems with constant boundary conditions have been studied extensively.
Comparatively less work has been carried out on the responses of these
systems to time dependent parameters. We report the modifications to the
probability density function of a two particle exclusion model in response
to a periodically changing perturbation to its boundary conditions. The
changes in the shape of the distribution as a function of the frequency of
the perturbation contains considerable structure. A dynamical phase transition in which the system response changes abruptly as a function of perturbation frequency was observed. We interpret this
structure to be a consequence of the existence of a typical time-scale associated with the dynamics of density shock profiles within the system.
\end{abstract}

\maketitle
\section{INTRODUCTION}
Driven diffusive systems have been of interest to a wide community of researchers since the first models were introduced \cite{Katz1,Katz2}. Mostly motivated by their ability to demonstrate the curious phenomena of non-equilibrium systems despite their theoretical simplicity, various different systems of such are proposed \cite{solversguide,rep1,rep2,rep3,rep4,rep5,schutz2part}. Asymmetric simple exclusion processes (ASEP) being one of the simplest of those, with particles interacting exclusively, hopping to both directions or in and out of the system with certain probability rates on a one dimensional lattice, can demonstrate the interesting theoretical phenomena such as phase separation \cite{Krug}, spontaneous symmetry breaking \cite{spontaneous}, phase coexistence \cite{spontaneous}, and shock formation \cite{shocks}. Furthermore, they can be utilised in modelling various real life problems such as transport of inter-cellular motor proteins \cite{solversguide}, traffic jams \cite{rep4}, surface growth \cite{surfacegrowth}.

One dimensional asymmetrical simple exclusion process (ASEP) systems with open boundaries are bounded to particle baths of constant densities at both ends, which can be modelled with constant boundary crossing rates for particles which enter or leave the system. Although considerable amount of work exists on the time-independent steady state properties, there are only a few examples which study the effect of applying time-dependent or oscillatory boundary rates to these systems.  
For instance Popkov {\em et al.} apply an on and off boundary condition to the single-species, semi-infinite ASEP, such that the oscillating exit probability rates can be thought as red and green traffic lights \cite{schutz}. A significance of their result which is relevant to our work is the observation that the density fluctuations propagate with a typical velocity into the lattice from the boundary. The fluctuation although weakening as it propagates, preserves its characteristic shape within the bulk. They observe the density response of the system has a sawtooth-like
characteristics with periodic pileups related to the red-green light periods of the system independent of the initial conditions. They also showed same behaviour exists in the hydrodynamic limit. In another work, Basu {\em et al.}, applied a sinusoidal drive to the boundaries of single-species simple exclusion process (SEP) and ASEP models, in which particles are allowed to move to the in both directions in symmetric and asymmetric rates correspondingly. They performed Fourier analysis of the response of both systems. They found that the structure functions have bimodality, which they claim, indicate the modes of transportation in diffusive systems \cite{indians}. 

In this present work, we carry out a Monte Carlo study of a two-species totally asymmetric simple exclusion process (TASEP) such that the boundary conditions (BC) are abruptly oscillating with relatively small amplitudes around a phase transition point.  We show that the system responds in qualitatively different forms, depending on the frequency of the perturbation. In particular, there appears to be a dynamical transition where the response of the system changes from that of a symmetric state to that of an  near-symmetric one abruptly, as a function of the frequency of the perturbation. This transition is independent of the size of the perturbation.

The phases in the phase diagram of the time-independent model were first reported by Evans {\em et al.} \cite{evans}. They identified through a mean field analysis, four different phases of the order parameter density, for all symmetric parameters of the two-species. One of the phases surprisingly display broken symmetry. Between the symmetric and asymmetric phases they report a tiny regime in which particle densities are low but not symmetric. (We will label three of the phases of interest to us as $LL$ [symmetric low density-low density], $HL$[the broken symmetry high density-low density], and $TR$ [tiny regime].) We will give a precise definition of the model in the next section. For our general discussion at this point, we demonstrate how the joint density
function $p(n_1,n_2)$ behaves near $TR$ as a function of the boundary exit rates $\beta_1$ and $\beta_2$ for the two types of particles. (Arndt {\em et al.} discuss the structure of these phases in detail in \cite{Arndt}.)
\begin{figure}[h]
\includegraphics[width=8.6cm]{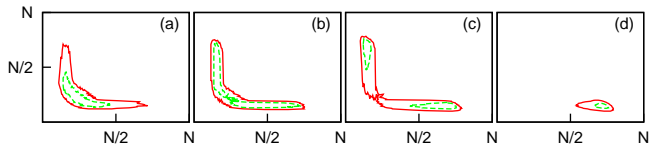}
\caption{\small (Color online) Joint probability densities $p(n_1, n_2)$ for number of particles for various paramaters under constant BC. For a lattice size of $N=200$, parameters for the plot $(a)$ are $\beta_1=\beta_2=0.285$, for the plot $(b)$ are $\beta_1=\beta_2=0.275$ and for the plot $(c)$ are $\beta_1=\beta_2=0.265$ and for the plot $(d)$ they are asymmetric as $\beta_1=0.265$ and $\beta_2=0.285$. Contours enclose approximately 75 and 25 percent of the total probability. For all of the plots all other transition rates are equal to $1$. \label{fig:combined}}
\end{figure}

$TR$ was shown to be a finite size effect by Erickson {\em et al.} \cite{pruessner}. Through a Monte Carlo analysis they showed that the size of this phase decays exponentially with respect to  lattice size. Detailed analysis of the joint density distributions of two types of particles for this regime reveals that the density is a superposition of ``shock profiles" along the length of the system \cite{schutz, Mallick, Arndt}. Each profile, which corresponds to a particular number of type I particles in the system, has an error-function like shape, whose midpoint is carrying out a random walk across the lattice \cite{Mallick}. The random walk is constrained when the shock approaches a boundary, if it gets too near the particle entry (exit) boundary, the increase (decrease) in the density near the boundary has a compensating effect on the position of the shock, pushing it away from the boundary. The entry and exit rates as a function of the position of the shock may then be interpreted as a ``force" on the shocks, with a corresponding ``potential", in which the random walk is carried out\cite{Arndt}. 

\begin{figure}[h]
\includegraphics[width=8.6cm]{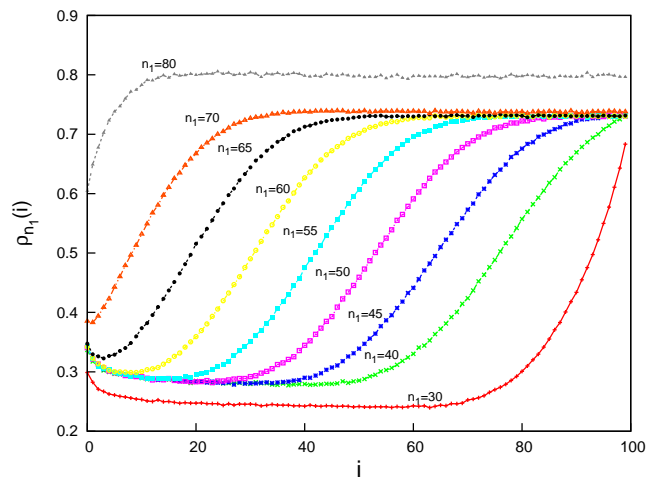}
\caption{\small (Color online) Shock profiles in a system of size $N=100$, for various occupation numbers $n_1$. The boundary rates are kept fixed at $\alpha_1=\alpha_2=1$ and $\beta=0.2675$. It can be observed that the profiles for $n_1<40$ and $n_1>70$ deviate in character from others due to boundary effect. \label{fig:equi_shocks}}
\end{figure}

Fig.~\ref{fig:equi_shocks} shows these profiles corresponding to several values of occupation of the lattice at time-independent steady state. The plots show the average density of first type of particles as a function of position, when the lattice contains a total of $n_1$ such particles with $n_1>n_2$. This last constraint limits the averaging to one leg of the boomerang-shaped probability density.
 
The discussion above points out two different features for the motion of the shock profile: The first corresponds to diffusive, damped motion in an effective potential. The second is the mechanism of application of an effective force through the manipulation of the boundary conditions, which will have a retarded effect dependent on the position in the lattice. We have looked into possibility of the production of interesting effects through the interplay of these two features. We investigate whether it is possible to force the shocks in the system by simply oscillating the boundary conditions. We observe significant frequency dependence which is unusual for a diffusive system.
We have also observed hysteresis in the density function of the system. Such behaviour was observed earlier in similar systems by Rakos  {\em et al.}\cite{Rakos}. Hysteresis in our model appears abruptly as perturbation frequency is decreased,
associated with a typical velocity in the system.
In the following sections we first introduce the model we are studying, and how we apply the oscillatory boundary conditions. We then move on to the discussion of the response of the system to the boundary conditions. 
\section{TASEP under Periodically Driven BC}
The system which is studied in this paper is TASEP on a finite, one-dimensional lattice with open boundaries and two species. Particles of type 1 (2) are allowed to enter the system from the left (right) with probability rate $\alpha_{1(2)}$, move only forward with probability rate $\gamma_{1(2)}$ if the following site is empty, and leave the system from the opposite end with probability rate $\beta_{1(2)}$. Different types of particles are allowed to switch places with rate $\delta$ when they come face to face. In our simulations all probability rates except the exit rates were taken to be equal to $1$. (These unitless quantities define a unitless time scale for the problem.) In comparison to the on and off exit rates of Popkov {\em et al.} \cite{schutz}, relatively small oscillations of the exit rates were applied to the system. We let the exit rate to oscillate around the $TR$ phase boundary point $\beta_o=0.275$ with an amplitude $\Delta\beta$:
\begin{eqnarray}
\beta_1 &=& \beta_0-\Delta\beta s(t) \\ 
\beta_2 &=& \beta_0+\Delta\beta s(t) \nonumber
\end{eqnarray}  
where $s(t) = \mbox{sgn} (\sin(2\pi t/\tau))$ for time $t$ within a period of oscillation $\tau$.

To study the system we use Kinetic Monte Carlo simulation \cite{kinetic}, with Poissonian time dynamics. We maintain a list of all possible events $e$ (possible particle jumps within the system and motion through
the boundaries) and rates $\omega_e$ associated with them. The total rate for any one of these events happening is then given by $\Omega = \sum_e \omega_e$. A random variable $\Delta t$ which corresponds to the time increment for the next event is then given by $\Delta t = -\log(r)/\Omega$ with $r$ a random number uniformly distributed between $0$ and $1$. If $\Delta t$ implies a time increase past the next BC change time $t_t$ given by eqn 1, no changes are made to the system and the time is set to $t_t$. Otherwise we select the
particular type of change $e$ that takes place at that time randomly, with the probability $\omega_e /\Omega$. The procedure described in this paragraph is then repeated, and statistical averages evaluated, weighing the influence of each state with $\Delta t$. To produce such random numbers we use the Mersenne Twister pseudo-random number generator \cite{Mersenne1, Mersenne2}.

In the simulations a Monte Carlo step (MCS) was taken to be $N \times N$ time increments. This definition of MCS may be associated with the maximum lattice transit time of the particles in the system. (A particle has to make $N$ jumps to transit the system. For a ``typical" distribution of particles, each jump will take $\mathcal{O}(1)$ time unit, and $\Delta t$ is $\mathcal{O}(1/N)$.) Note that we also have a continuous time variable $t$ associated with time increments $\Delta t$, but large MCS is used to ensure good statistics. In each simulation, averages are calculated over $10^5$ MCS. Period dependent averages are calculated by obtaining time dependent averages within each period and averaging over the periods.   
\section{Variations in the character of Frequency Dependence}
We change the boundary conditions in a way to break the symmetry between the two types of particles. We choose $TR$ as the unperturbed state which is associated with the presence of shock fronts in the widest region. Note that at very high frequencies, one obtains the unperturbed state, while at very low frequencies, the system moves from one constant asymmetric BC state to the other. To discuss these varying responses we focus on the joint density distributions for some values of period of oscillations. Depending on the frequency of oscillation, we observe very different types of responses.\\

\begin{figure}[h!]
\includegraphics[width=8.9cm]{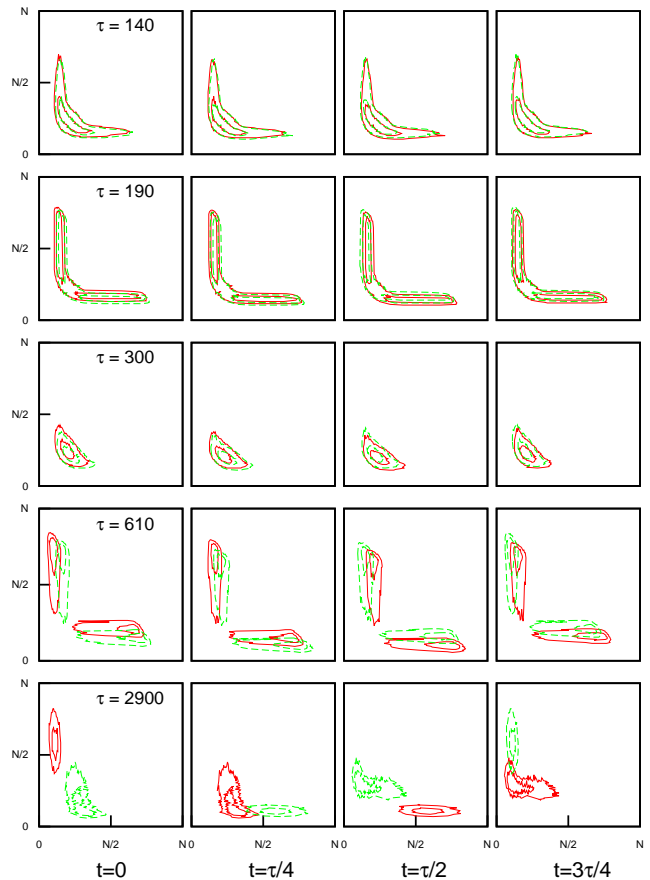}
\caption{\small (Color online) Time dependence of the joint density distribution $\rho(n_1,n_2)$ corresponding to the significant points of the Fig.~\ref{fig:gezinti_harfli}. In each column, the density at the indicated time as well as at the next quarter cycle (dashed lines) are shown to display the change. \label{fig:contours_bitisik}}
\end{figure}

Fig.~\ref{fig:contours_bitisik} displays the change in the joint distribution function for a system of size $N=200$ and $\Delta\beta=0.1$, for various frequencies. Note that the boomerang shaped profile (similar to those in Fig.~\ref{fig:combined}) disappears and re-appears as a function of frequency. At high frequencies of oscillation (low values of $\tau$) the density distribution preserves the boomerang shaped nature (similar to those in Fig.~\ref{fig:combined}) but the distribution tends to move as a whole in response to the changing boundary condition. We use the terminology ``near-symmetric"
states in association with such density functions, which although not preserving perfect symmetry between the two types of particles, maintain a shape which is a perturbation of the symmetric 
time-independent version.  This shape itself varies as the oscillation frequency is changed, resembling the time independent density distributions for different values of the parameter $\beta$ in Fig.~\ref{fig:combined}. (All figures display results for system size $N=200$, except where $N$ dependence is stated.) However note that in order to observe a time-independent distribution similar to that for $\tau=300$ in Fig.~\ref{fig:contours_bitisik}, one would need to go deeper into the $LL$ phase than the range of  parameters used in oscillating BC (See Fig.~\ref{fig:combined}.~a). This is an indication of the resonance-like behaviour in the system; driving the density fluctuations much higher than the values one can obtain from the static values in the same range. Although the joint density is confined to a very small range $\tau=300$, smaller and longer values of $\tau$ result in densities which are still boomerang shaped. 
 
\begin{figure}[h!]
\includegraphics[width=8.9cm]{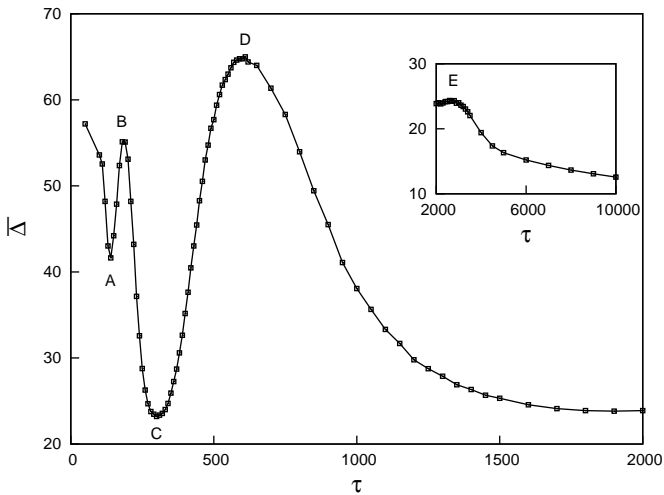}
\caption{\small Average density spread ($\bar{\Delta}$) graph with respect to different period values. Inset shows the average spread for higher values of period. Interesting points are labeled with letters. Periods corresponding to those letters are: $\tau=140$ for $A$, $\tau=190$ for $B$, $\tau=300$ for $C$, $\tau=610$ for $D$ and $\tau=2900$ for $E$. Density distribution for these points are give in Fig.~\ref{fig:contours_bitisik}. \label{fig:gezinti_harfli}}
\end{figure}

To quantify this behaviour we introduce a parameter, which we call ``spread'', defined as follows: We calculate the averages below at 100 time values $t_i = i \tau/100$ within each period $\tau$: 
\begin{eqnarray}
{\langle n_1^m\rangle}_{t_i} = \sum_{n_1,n_2} n_1^m~p(n_1,n_2,t_i) \nonumber \\
{\langle n_2^m\rangle}_{t_i} = \sum_{n_1,n_2} n_2^m~p(n_1,n_2,t_i) \nonumber \\
\Delta_1^2(t_i) = {\langle n_1^2\rangle}_{t_i} -{{\langle n_1\rangle}^2_{t_i}} \nonumber \\
\Delta_2^2(t_i) = {\langle n_2^2\rangle}_{t_i} -{{\langle n_2\rangle}^2_{t_i}} \nonumber
\end{eqnarray}
Then average spread is:
\begin{equation}
\bar{\Delta} = \sqrt{\frac{\sum_i(\Delta_1^2(t_i)+\Delta_2^2(t_i))}{100}} \nonumber
\end{equation}
This is then an average of the fluctuation in the number density during a period of oscillation.

Fig.~\ref{fig:gezinti_harfli} is a plot of this parameter as a function of oscillation period and indicates that system is going through a resonance-like behaviour at various frequencies. Extrema on this plot are identified with letters $A-E$ and correspond to the distributions in Fig.~\ref{fig:contours_bitisik}. For instance for $\tau=140$ (point A on Fig.~\ref{fig:gezinti_harfli}), density is mainly distributed around the $LL$ region with some tails into symmetry broken states. When $\tau=190$ (point B) joint density closely resembles the equilibrium density. The minimum at $C$ corresponds to the very compact distribution mentioned above. On the other hand, for low frequencies, {\em e.g.} for when $\tau=2900$ (point E), the system is in the broken symmetry state at all times within the period. The appearance of large scale hysteresis is apparent in this case. We discuss below the abrupt appearance of this hysteresis effect.
For even lower frequencies, the system is driven deeper into the symmetry broken phase at each half cycle, resulting in an even smaller spread as the inset to Fig.~\ref{fig:gezinti_harfli} displays. 

\begin{figure}[h!]
\includegraphics[width=8.9cm]{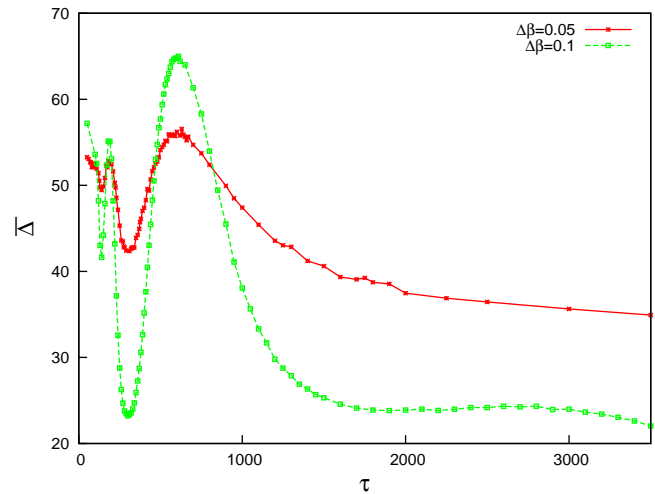}
\caption{\small (Color online) Average density spread ($\bar{\Delta}$) responses of the system for different magnitudes of perturbation, $\Delta \beta = 0.05$ and $\Delta \beta = 0.1$. Note that the $\tau$ values for the extrema of the response is independent of the size of the periodic drive. \label{fig:amplitude_gezinti}}
\end{figure}

\begin{figure}[h!]
\includegraphics[width=8.9cm]{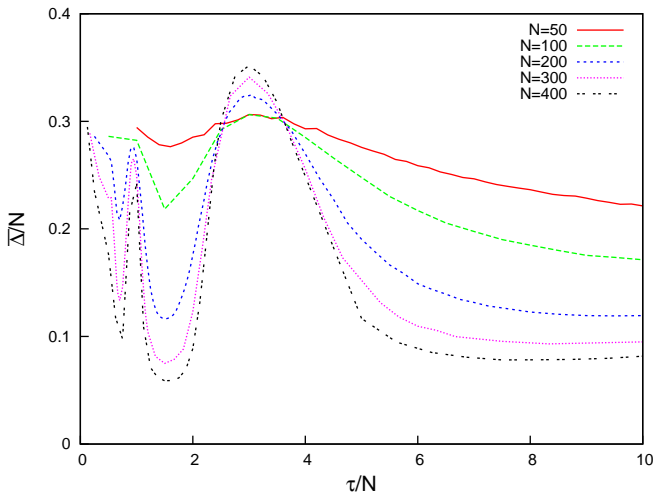}
\caption{\small (Color online) Average density spread ($\bar{\Delta}$) with respect to oscillation period $\tau$ for various values of lattice sizes. Both axes are scaled by N. \label{fig:coklu_gezinti}}
\end{figure}

The effect of the amplitude $\Delta \beta$ of the perturbation on the spread parameter is shown in Fig.~\ref{fig:amplitude_gezinti}. The structure of the response is preserved, but the magnitude dependence is apparent. Smaller perturbation leads to smaller variation in spread at higher frequencies. However, the spread
diminishes less slowly at longer periods because it takes a longer time to push the system into the asymmetric phase with a smaller perturbation. Fig.~\ref{fig:coklu_gezinti} displays the effect of the system size on the
response. Existence of a ``typical velocity" in the system would lead to an expectation of scaling of all characteristic time constants by $N$. Fig.~\ref{fig:coklu_gezinti} indicates that this is indeed the case. However characteristic times (such as response extrema) are not simply related to one another, indicating that the size of the boundary regions (which should be excluded from $N$) may be different for mechanisms which are responsible for various extrema. \\

\begin{figure}[h!]
\includegraphics[width=8.9cm]{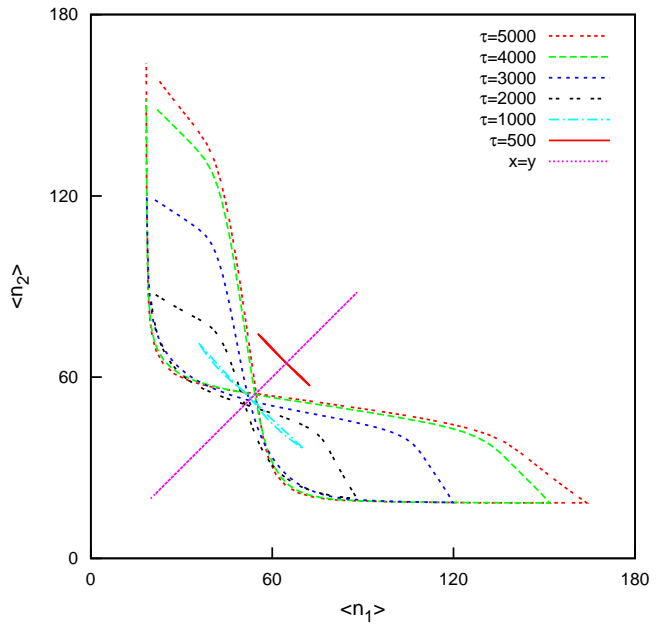}
\caption{\small (Color online) The trajectory for ${\langle n_2 \rangle}_{t_i}$ vs ${\langle n_1 \rangle}_{t_i}$ for values of $t_i$ within a period. The trajectories collapse to their limiting forms for $\tau \leq 5$ and $\tau \geq 5000$. Above $\tau =5000$ hysteresis area saturates, all overlap with $\tau =5000$.   \label{fig:kelebekler}}
\end{figure}

We have looked at the hysteresis in the average values ${\langle n_1 \rangle}_t$ vs ${\langle n_2\rangle}_t$ of the joint probability distribution function $p(n_1,n_2,t)$ in some detail. Fig.~\ref{fig:kelebekler} displays this effect for various values of the oscillation period. 

\begin{figure}[h!]
\includegraphics[width=8.9cm]{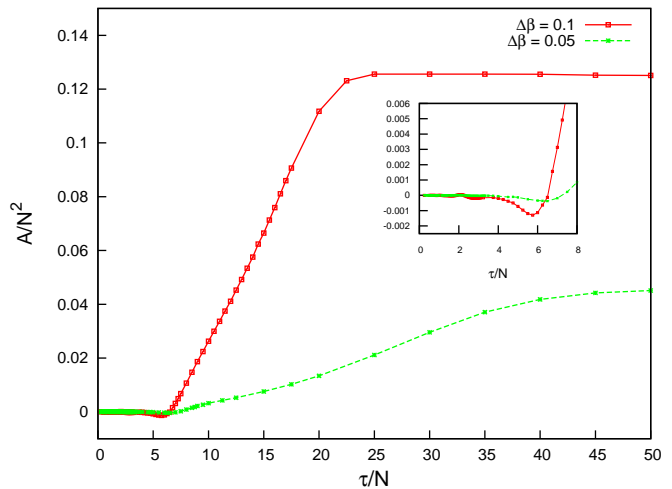}
\caption{\small (Color online) Hysteresis area for two different perturbation amplitudes, $\Delta \beta =0.05$ and  $\Delta \beta =0.1$ and $N=200$. The inset shows the detail near the break-away point. \label{fig:hist_area}}
\end{figure}

We calculate the area of the hysteresis curve 
\begin{equation}
A=\sum_{t_i} {\langle n_2\rangle}_{t_i} \Delta {\langle n_1\rangle}_{t_i} \nonumber \\
\end{equation}  
where $ \Delta {\langle n_1\rangle}_{t_i} = {\langle n_1\rangle}_{t_i}-{\langle n_1\rangle}_{t_{i-1}}$. Here the summation is over one lobe of the ($8$-shaped) hysteresis loop. Fig.~\ref{fig:hist_area} displays the result. Although some amount of hysteresis (not visible at the scale of Fig.~\ref{fig:hist_area}) exists at all frequencies, we find that a large-scale hysteresis starts at $\tau \sim 5N$, independent of $\Delta\beta$ or $N$. This may be interpreted as the onset of large scale motion of the probability density associated with symmetry broken phase. We then identify the value $N/\tau\sim 0.2$ as a typical velocity in the system. Hysteresis is not present when changes to the system are faster than that implied by this characteristic velocity. The inset to Fig.~\ref{fig:hist_area} shows that there is some structure associated with the break away point of the hysteresis magnitude. We identify this point as a dynamical phase transition point as a function of frequency. It is interesting to note that the values for which $\tau/N < 5$ correspond to the range in Fig.~\ref{fig:coklu_gezinti} where $\bar{\Delta}$ displays richer structure.

\section{Pulse Response}
To better understand the nature of the frequency dependence of the system, we further study the ``pulse response'': We have applied a constant perturbation, only to the exit rate of the first type of particles, $\beta_1=0.535$, for a duration of $\Delta t=100$ over a period of $\tau=10000$ with a repetition for $10^7$ MCS. When time reaches the end of the period, the system is relaxed to a near time-independent steady state. We have thus obtained the time-dependent shock profiles and average occupation values, which again show surprising oscillatory behaviour.

\begin{figure}[h]
\includegraphics[width=8.6cm]{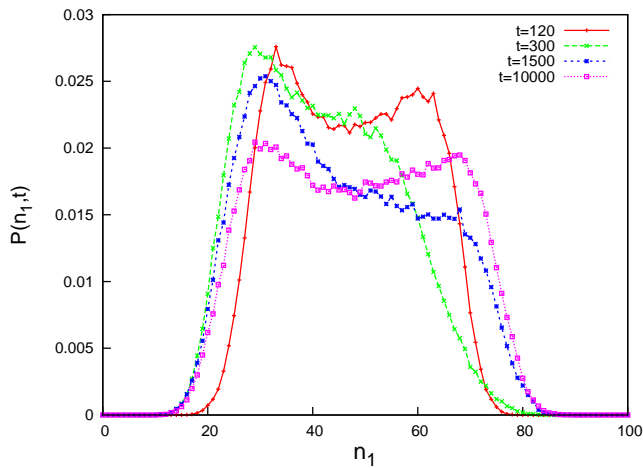}
\caption{\small (Color online) $P(n_1,t)$ for various values of $t$. $P(n_1,10000)$ is a near steady-state distribution. \label{fig:prob_dist}}
\end{figure}

\begin{figure}[h]
\includegraphics[width=8.6cm]{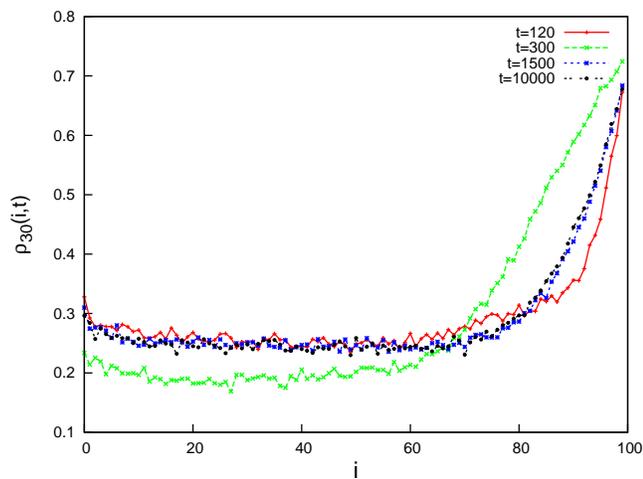}
\caption{\small (Color online) Shock densities corresponding to $n_1=30$ for various values of $t$.\label{fig:kucukn}}
\end{figure}

\begin{figure}[h]
\includegraphics[width=8.6cm]{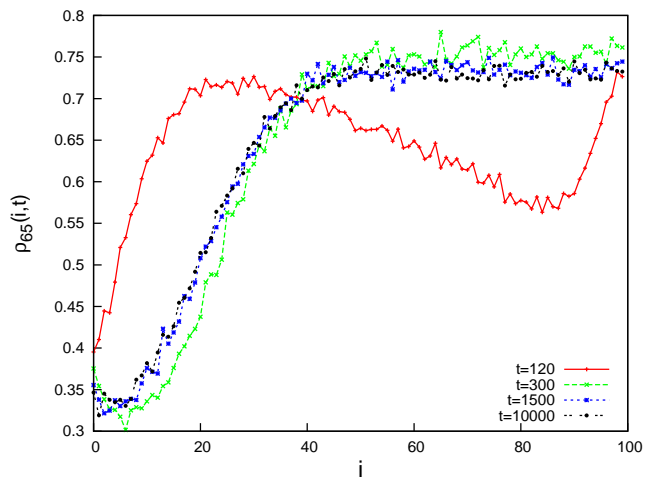}
\caption{\small (Color online) Shock densities corresponding to $n_1=65$ for various values of $t$. At early times, the profile has a distortion which indicates loss of particles on the right hand (exit) side due to the pulse. The entry side resembles a steady-state profile for a larger $n_1$ in Fig.\ref{fig:equi_shocks}. The perturbation pulse acts to shift the densities for large $n_1$ shocks to lower $n_1$ values. \label{fig:buyukn}}
\end{figure}

Fig.~\ref{fig:prob_dist} shows the variation of the occupation number probability of first type of particles in one leg of the boomerang-shaped probability density:
\begin{equation}
P(n_1,t) = A \sum_{n_2=0}^{n_2=n_1-1} p(n_1,n_2,t)\\ \nonumber
\end{equation}
where $A$ is a normalization constant. Note that, probability is reduced at early times for smaller and larger values of $n_1$. Figures \ref{fig:kucukn} and \ref{fig:buyukn} show the shock 
profiles at various times after the pulse. It can be observed that the profiles for small $n_1$ show less of a distortion compared to those for larger $n_1$. The profiles for larger $n_1$ are distorted due to the exit of particles during the pulse. The change in $P(n_1,t)$ for small times is then due to two different mechanisms: The small $n_1$ shocks simply leave the system during the pulse, while the large $n_1$ shocks are deformed into smaller $n_1$ forms. The recovery of the system from these two effects seem to be qualitatively different. The statistics for large $n_1$ shock recover exponentially with a dynamics consistent with a diffusive system. Recovery of small $n_1$ shock statistics seem to be a contribution of multiple effects, resulting in an oscillatory damping.
 
\begin{figure}[h]
\includegraphics[width=8.6cm]{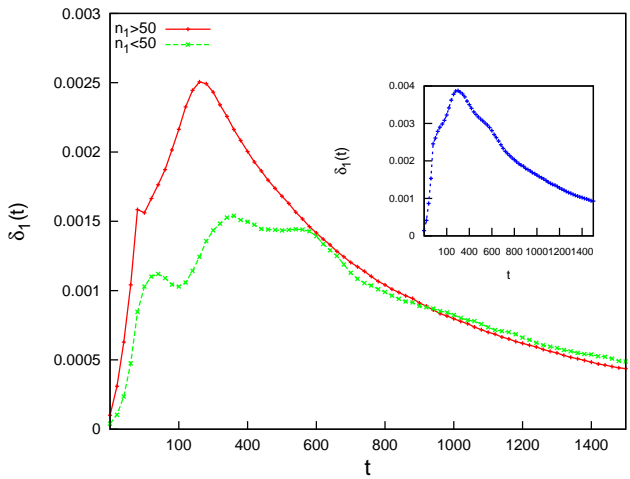}
\caption{\small (Color online) Relaxation of the deviation $\delta_1(t)$ for smaller and larger values of $n_1$ for $N=100$. Inset shows $\delta_1(t)$ for all values of $n_1$. \label{fig:diff_fr}}
\end{figure}

The inset to the Fig.~\ref{fig:diff_fr} displays the relation of $n_1$ to its steady state values as a function of time:
\begin{equation}
\delta_1(t)=\sum_{n_1}(P(n_1,t)-P(n_1,\infty))^2.
\end{equation}
To separate the two mechanisms discussed above, Fig.~\ref{fig:diff_fr} shows the contributions to this summation for values of $n_1<N/2$ and $n_1>N/2$. Note that for both cases deviation from the steady state increases for a period of time even after the perturbation pulse has ended. However, smaller $n_1$ statistics relax to the steady state with shorter time scale oscillations suggesting that the process may be associated with boundary events rather than the bulk. The oscillatory nature of the relaxation to steady state is also apparent in Fig.~\ref{fig:diff_fr}. This unusual behaviour forms the basis of the different type of response we report for the sinusoidal drive. 

One does not expect to find an oscillatory response in a diffusive system. The effect seems to be a superposition of a number of recovery processes with different time scales dominated by the statistics of states with smaller number of particles. The time scale of the oscillations is consistent with our report of $N/\tau \sim 0.2$ for the sinusoidal drive. More work may be necessary to identify the details of the mechanisms involved in this interesting phenomena.
 
\section{CONCLUSION} 
We report the response of the TASEP model as a function of the perturbation
frequency of the boundary condition. We find that the response is qualitatively different for various ranges of the perturbing frequency. One type of change involves significant
modifications in the shape of the joint distribution function, which alternates between compact
and extended forms. Variation of this behaviour as a function of frequency
contains considerable structure which does not depend on the size of the periodic drive, and scales with the size of the system. This  implies  that  the  response  is  associated  with  motion of features through the system, in the form of shock fronts. A second type of change that was observed is the abrupt appearance of the hysteresis as the frequency of the perturbation is lowered.  This also indicates a velocity threshold under which the the density distribution cycles from one phase (associated with that particular value of the BC) to the other, with significant changes during the cycle. We identify a characteristic velocity value of $\sim 0.2$ lattice sites per unit time. Higher frequencies correspond to near-symmetric states where the probability distribution moves more or less rigidly during the cycle,
albeit with a probability density profile which changes appreciably as a function of frequency.

We have reported the response of the system at a phase transition point, which we thought would be most interesting. Analysis of other special points on the phase diagram could also shed light on the dynamical mechanisms of interest in this system.

Authors acknowledge support from Turkish Academy of Sciences (TUBA).

\end{document}